# Effect of nitrogen doping and pressure on the stability of LuH$_3$

Yang Sun[1,2*], Feng Zhang[2], Shunqing Wu[1*], Vladimir Antropov[2*], and Kai-Ming Ho[2]

[1]Department of Physics, Xiamen University, Xiamen 361005, China

[2]Department of Physics, Iowa State University, Ames, IA 50011, USA

**Abstract**

The report on the near-ambient superconductivity in a nitrogen-doped lutetium hydride has stimulated great interest in this material (Dasenbrock-Gammon et al. 2023). While its superconductivity is still a subject of debate, the structure of the claimed cubic phase remains uncertain. In this work, we study the effect of nitrogen doping and pressure on the energetic and dynamic stability of cubic LuH$_3$. Our findings indicate that both pressure and nitrogen doping can enhance the stability of the cubic LuH$_3$ phase. We propose a Lu$_8$H$_{21}$N structure that exhibits stable phonon, reasonable thermodynamic stability at 1 GPa, and a similar XRD pattern to the experimental data. However, we do not observe electron-phonon coupling in the zone-center phonon modes of these crystal structures.

[*]Email: yangsun@xmu.edu.cn (Y.S); wsq@xmu.edu.cn (S.W.); antropov@iastate.edu (V.A.)

The search for room-temperature superconductors under ambient conditions is a huge scientific challenge that can lead to many applications. More than 50 years ago Ashcroft and Ginzburg already discussed the possibility of having such superconductivity in the metallic phase of hydrogen [1,2]. Recently it became possible to predict and observe new hydrogen-rich metallic phases with such superconductivity [3]. High-pressure experiments have had great success in discovering many high-$T_c$ hydrides, such as $H_3S$ with $T_c$ above 200 K under 200 GPa [4], $LaH_{10}$ with $T_c \sim 250K$ at 170GPa [5,6], $ThH_{10}$ with $T_c \sim 160K$ at ~170GPa [7], $CeH_{10}$ with $T_c \sim 115K$ at 95GPa [8], $YH_6$ with $T_c \sim 224K$ at ~166GPa [9,10], $YH_9$ with $T_c \sim 262K$ at ~180GPa [10,11], $CaH_6$ with $T_c \sim 215K$ at 172GPa [12]. However, such high pressures are still too extreme for practical applications. Therefore, doping or substituting hydrogen with other light elements such as B, C, and N has been proposed to lower the required high pressure [13–22].

Recently, Dasenbrock-Gammon *et al.* reported experimental evidence of superconductivity in N-doped Lutetium hydride near ambient conditions [23]. The material was synthesized at 2 GPa and claimed to achieve the $T_c$ of 274 K at 1 GPa. However, the superconducting behavior of this phase is being questioned [24,25], and the crystal structure of the synthesized phase is unclear. Dasenbrock-Gammon *et al.* described the structure as a cubic $LuH_{3-\delta}N_\varepsilon$, where δ and ε are unknown values that account for the hydrogen vacancy defect and nitrogen contents, respectively. As the X-ray diffraction pattern is not sensitive to hydrogen and nitrogen, only the lattice parameter for the Lu sublattice was solved with a cubic structure with a=5.0289(4) Å. Elemental analysis revealed a weight percent of 0.8-0.9% N in the materials. However, the cubic $LuH_3$ is known to be unstable under ambient conditions. Therefore, the unknown crystal structure makes the study of superconductivity in this compound difficult. In this work, we use first-principles calculations to investigate the effects of pressure and N doping on the thermodynamic and dynamical stability of cubic $LuH_3$. Our aim is to find a stable structure model that can explain the experimental data reported in Dasenbrock-Gammon et al.

*Methods.* Density functional theory (DFT) calculations were carried out using the projector augmented wave (PAW) method [26] implemented in the VASP code [27,28]. PAW potentials with valence electronic configurations *$6s^2 5d^1 5p^6$*, *$1s^1$*, and *$2s^2 2p^3$* were used for Lu, H, and N atoms, respectively. The exchange and correlation energy was treated with the generalized gradient approximation (GGA) and parameterized by the Perdew-Burke-Ernzerhof formula (PBE) [29]. A plane-wave basis was used with a kinetic energy cutoff of 520 eV, and the convergence criterion for the total energy was set to $10^{-8}$ eV. The Γ-centered Monkhorst-Pack grid was adopted for Brillouin zone sampling with a spacing of $2\pi \times 0.025$ Å$^{-1}$. The convex hull in this work is based on static enthalpy calculations. The reference phases to compute the convex hull are consistent with Material Project and OQMD databases, which use the $P6_3/mmc$ phase for Lu, $P6_3/mmc$ phase for H$_2$ and $Pa\bar{3}$ phase for N$_2$. The full Brillouin phonon spectrums were computed by the finite displacement method implemented in the Phonopy code [30] using $4 \times 4 \times 4$ supercell for LuH$_3$ (256 atoms) and $2 \times 2 \times 2$ supercell for Lu$_8$H$_{21}$N$_1$ (240 atoms). The Gaussian smearing scheme is performed in the calculations. We tested Methfessel-Paxton and Fermi-Dirac smearing schemes and obtained results consistent with each other.

The zone-center electron-phonon coupling strength ($\lambda_\Gamma$) was calculated by the frozen-phonon method, which is the relative difference between the screened ($\omega$) and unscreened ($\tilde{\omega}$) zone-center phonon frequencies [31] as

$$\lambda_\Gamma = \frac{\tilde{\omega}^2 - \omega^2}{4\omega^2}. \tag{1}$$

The screened phonon frequency was computed by fully self-consistent field (SCF) calculations in the displaced atomic configurations using the tetrahedron method with Blöchl corrections. To compute the unscreened phonon frequency $\tilde{\omega}$, an SCF calculation with the tetrahedron method was first performed in the equilibrium configuration, followed by the SCF calculations with the displaced atoms, but with partial occupations fixed as the one in the equilibrium configuration. The detailed workflow of this method can be found in [31].

The binary LuN [32], LuH$_2$ [33] and LuH$_3$ [34] phases have been synthesized and documented in the experimental ICSD database, but no stable ternary Lu-H-N phase has been reported yet. Computational databases, such as Material Project [35], Atomly [36] and OQMD [37], yield consistent results. Lu-H-N system at 1 GPa shows a relatively simple phase diagram in which only LuN, LuH$_2$, and LuH$_3$ are stable phases (Supplementary Material Fig. S1 [38]). Both LuN and LuH$_2$ have the Fm$\bar{3}$m symmetry, with Lu occupying the fcc sublattice. In LuN, N occupies the octahedral site, while in LuH$_2$, H occupies the tetrahedral site. The stable LuH$_3$ phase exhibits a rhombohedral structure (noted as r-LuH$_3$) with space group P$\bar{3}$c1, where Lu atoms pack in an hcp sublattice. The cubic LuH$_3$ (noted as c-LuH$_3$) shows a relatively high enthalpy, with 85 meV/atom higher than r-LuH$_3$ at 1 GPa (Table 1).

**Table 1.** Enthalpy above the static convex hull of LuNH phases at 1GPa.

| System | $H_{hull}$ (meV/atom) |
|---|---|
| $LuH_2$ | 0 |
| Rhombohedral $LuH_3$ | 0 |
| Cubic $LuH_3$ | 85 |
| $Lu_4H_{11}N$, N at octahedral site | 278 |
| $Lu_4H_{11}N$, N at tetrahedral site | 134 |
| $Lu_8H_{23}N$, N at octahedral site | 150 |
| $Lu_8H_{23}N$, N at tetrahedral site | 130 |
| $Lu_8H_{21}N$, N at tetrahedral site | 31 |

We first investigate the pressure effect on the stability of c-$LuH_3$. The phonon spectra of c-$LuH_3$ at various pressures are shown in Fig. 1. At pressures ranging in 1-5 GPa, the phonon modes show strong instability, which is mainly caused by the octahedral H atoms. As the pressure increases to ~20 GPa, the imaginary phonon modes are significantly reduced. At 25GPa, c-$LuH_3$ does not show any imaginary phonon modes. The phonon density of states shows that the H atoms at the tetrahedral sites have much higher frequencies than those at octahedral sites. At 25 GPa, the enthalpy of c-$LuH_3$ is only 12 meV/atom higher than that of r-$LuH_3$, which is significantly reduced compared to 85 meV/atom at 1 GPa. These results suggest that high-pressure conditions can stabilize c-$LuH_3$ both energetically and mechanically. We note that anharmonicity and quantum effects can affect the dynamical stability, possibly reducing the required pressure to stabilize the cubic phase [39].

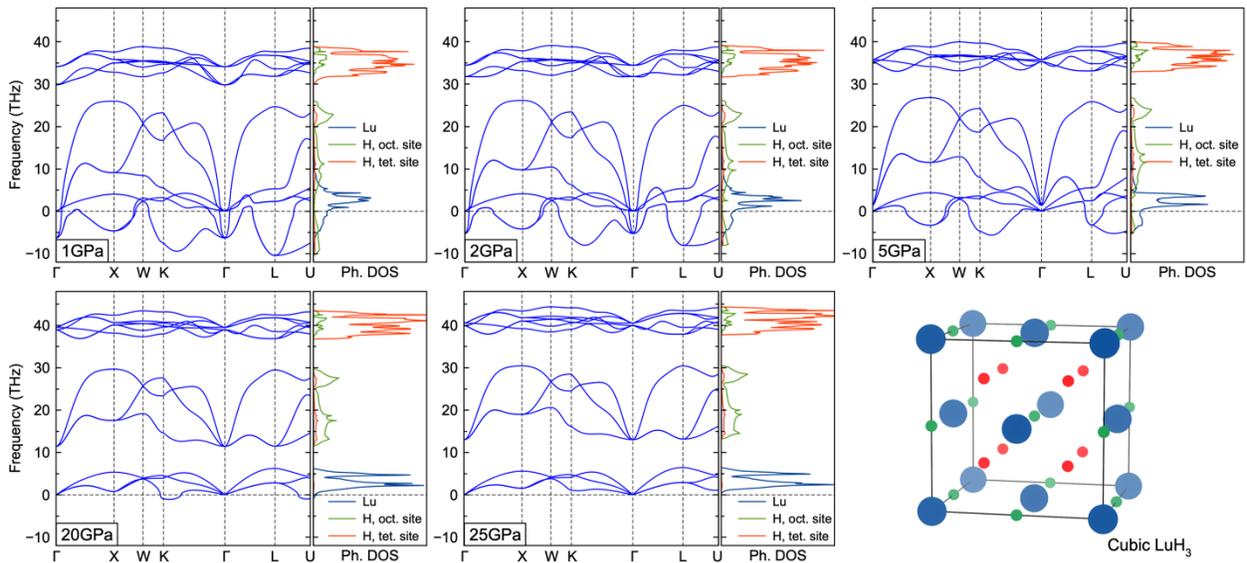

**Fig. 1** The phonon spectrum and phonon density of states (Ph. DOS) for cubic LuH$_3$ as a function of pressures. The imaginary modes disappear when the pressure is higher than ~25 GPa. In the atomic structure of cubic LuH$_3$, red shows H at the tetrahedral sites, while green shows H at the octahedral sites. Blue is Lu atoms.

*N doping effect on cubic LuH$_3$.* Although pressure can stabilize LuH$_3$, the condition of 25 GPa is too high compared to the near-ambient condition reported in Dasenbrock-Gammon *et al.* [23]. Hence, we dope the c-LuH$_3$ with N to study its effect on the stability at 1 GPa. We first use a conventional c-LuH$_3$ cell (4 f.u.) and substitute one H at an octahedral or tetrahedral site with an N, resulting in two inequivalent Lu$_4$H$_{11}$N structures. At 1 GPa, the enthalpies of both phases are well above the convex hull (Table 1), even higher than c-LuH$_3$. The enthalpy of the structure with N at the tetrahedral site is lower than the one with N at the octahedral site, indicating that N prefers the tetrahedral sites of c-LuH$_3$. N-doping at the octahedral site did not significantly alter the structure. However, doping at the tetrahedral site causes a change in the local H positions, as shown in Fig. 2(a). N repels the surrounding H atoms, leading to the formation of short H-H bonds and a local H tetrahedron. The phonon instability is contributed by the H atoms that form the local tetrahedron, as shown in Supplementary Material Fig. S2 [38]. The N content in Lu$_4$H$_{11}$N is 1.9 wt.%, which is twice larger than the ones (0.8-0.9wt.%) measured by Dasenbrock-Gammon *et al.* [23]. To be closer to the experimental condition, we use a $\sqrt{2} \times \sqrt{2} \times 1$ supercell (8 f.u.) and replace one H with N to reduce the doping concentration, yielding Lu$_8$H$_{23}$N. As shown in Table. 1, Lu$_8$H$_{23}$N with N at the tetrahedral site still shows a lower enthalpy than that at the octahedral site. Thus, if N is doped into the c-LuH$_3$, it is expected to occupy the tetrahedral sites. Both Lu$_4$H$_{11}$N and Lu$_8$H$_{23}$N show strong imaginary phonons in Fig. 2 and higher relative enthalpy above the convex hull compared to c-LuH$_3$. Therefore, these phases are still unstable and cannot be used to model the experimentally claimed LuH$_{3-\delta}$N$_\varepsilon$ phase [23].

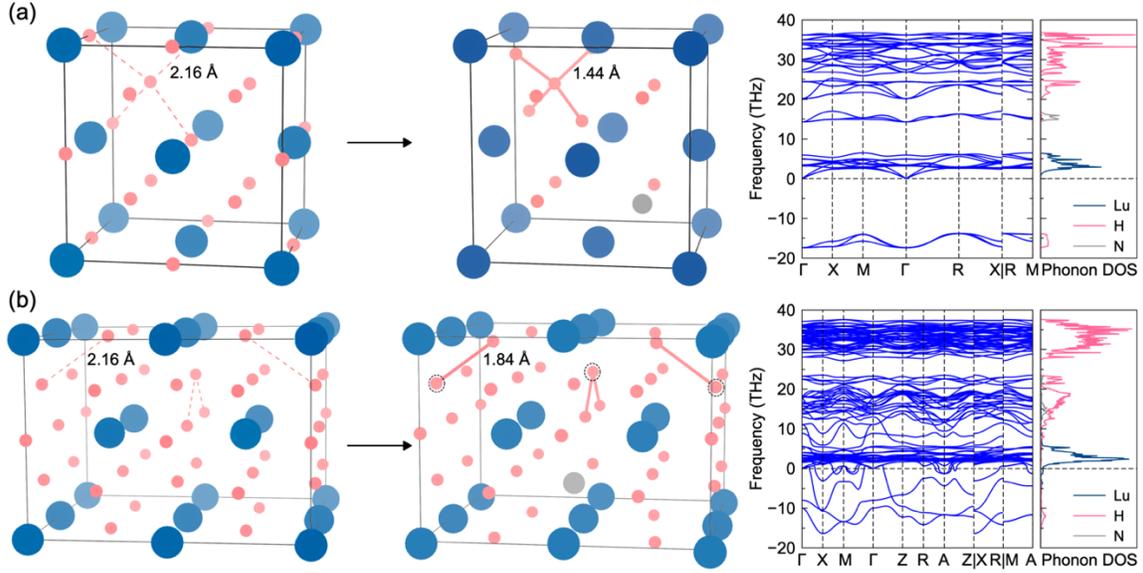

**Fig. 2** The structural change in (a) $Lu_4H_{11}N$ and (b) $Lu_8H_{23}N$ due to the N doping at the tetrahedral site at 1 GPa. The connected H-H bonds are the shortened bonds caused by the N substitution. The circled sites indicate the vacancies. The right panels show the phonon spectrum and density of states of doped structures.

By analyzing the phonon instability at the zone center in $Lu_8H_{23}N$, we find it is mainly caused by H atoms with shortened H bonds. To eliminate these instabilities, we introduce vacancies at the H sites that are closely related to these short bonds (circled sites in Fig. 2(b)). This generates a model with the composition of $Lu_8H_{21}N$. After relaxing this model, we find the phonon instability disappears, as shown in Fig. 3. Moreover, by introducing the defects, the enthalpy above the convex hull decreases to 31 meV/atom (Table 1). Thus, these defects significantly improve both thermodynamic and dynamic stability. As this structure contains 14 tetrahedral sites, the N atom can also replace other tetrahedral sites equivalently. It could form solid solutions at finite temperatures, introducing configurational entropy to further lower the free energy. If we estimate the configurational entropy with the ideal solid solution model as $S_{conf} = -k_B[x\ln x + (1-x)\ln(1-x)]$ where $x$ accounts for N concentration in the tetrahedral sites as $x = \frac{1}{14}$, the free energy above the convex hull, i.e., $\Delta G = \Delta H - TS$, is ~24 meV/atom at 294 K, which is within the thermal fluctuation at room temperature. Therefore, from the perspective of energetics and dynamical stabilities, this $Lu_8H_{21}N$ structure can exist under near-ambient conditions. We note that the vibrational and anharmonic effects at finite temperatures may also change the energetic stability, which is neglected in the present work.

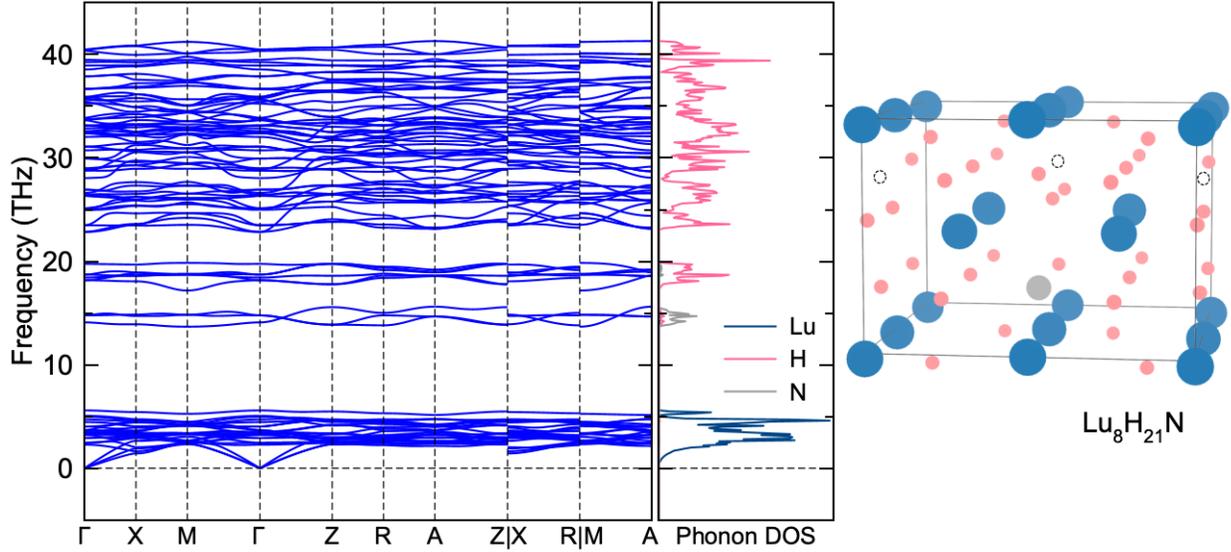

**Fig. 3** The phonon spectrum and atomic structure of $Lu_8H_{21}N$ at 1 GPa. The circled sites indicate the vacancies.

*Comparison of XRD patterns.* We compare the XRD of $Lu_8H_{21}N$ structures with the experimental data [23] in Fig. 4. The peak of $Lu_8H_{21}N$ XRD shows good agreement with experimental XRD without any refinement, although some differences exist. Firstly, it shows small peak splitting at $2\theta = 35°$, $51°$ and $61°$. These splittings are relatively small, even less than the FWHM of corresponding experimental peaks. They are mainly due to the change of lattice symmetry. With full DFT relaxation, the $Lu_8H_{21}N$ model is a tetragonal phase with space group $P\bar{4}m2$ (see crystallographic data in Supplementary Material Table S1 [38]). The a/c ratio is 1.406, which deviates from the original value of $\sqrt{2}$ ratio in the supercell of the cubic lattice. However, if it forms a solid solution at a finite temperature and N is randomly distributed to the tetrahedral site in the solid solution, the phase should recover the cubic lattice. Secondly, there is a systematical left shift of the peak positions of the fully relaxed structure compared to the experimental data. This is due to the lattice parameter difference. The XRD can be solved by the cubic lattice with $a = 5.0289$ Å, while the relaxed cell has $c = 5.1051$ Å. This can be attributed to the fact that PBE functional can sometimes overestimate the cell volume compared to the experimental measurement [40]. If relaxing the structure by LDA functional, the lattice parameter becomes 5.0115 Å (see Supplementary Material Table S2 [38]). This is smaller than the experimental value and caused a right shift of the XRD peaks, as shown in Fig. 4. To eliminate the effect of these two factors, we change the $Lu_8H_{21}N$ lattice to experimental one $a = b = 5.0289 \times \sqrt{2}$ Å and $c = 5.0289$ Å, while keeping atom positions fixed. This leads to an XRD that matches the main peaks of the experimental data well. With the experimental lattice the $Lu_8H_{21}N$ phase maintains the dynamical stability (see Supplementary Material Fig. S3 [38]). Considering the significantly lower enthalpy of the $Lu_8H_{21}N$ phase in comparison to other N-doped phases, it can be inferred that this phase is at a deep local minimum on the potential energy surface.

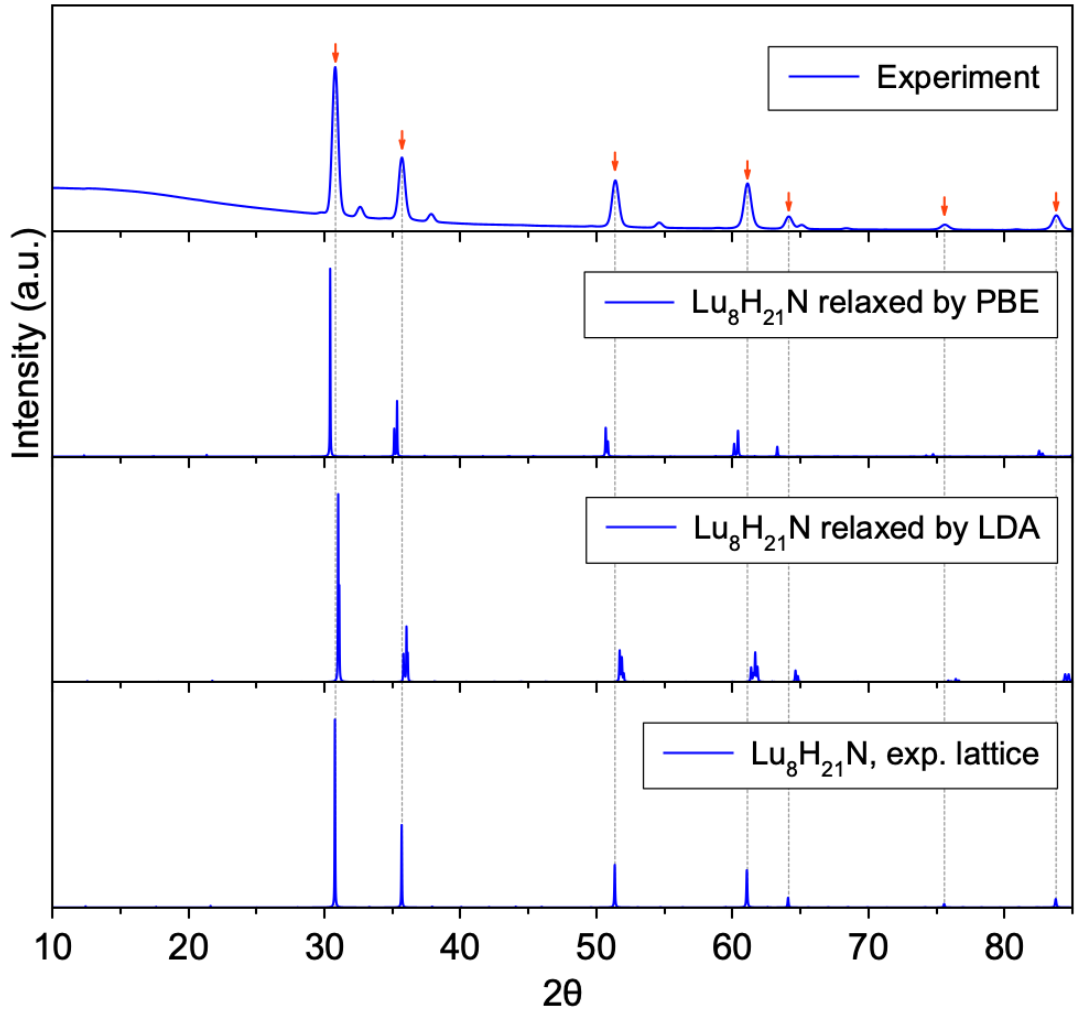

**Fig. 4** Comparison of XRD between experimental one and $Lu_8H_{21}N$ model with different lattices. The fully relaxed lattices by PBE and LDA functionals are shown in Supplementary Material Table S1 and S2 [38], respectively. The experimental lattice uses $a = b = 5.0289 \times \sqrt{2}$ Å and $c = 5.0289$ Å. Red arrows indicate the indexed peaks, which are also marked by dashed lines to guide the eye. The remaining peak can be explained by $LuN+Lu_2O_3$ as described in [23].

*Zone-center electron-phonon coupling strength.* We have demonstrated that both high pressure and N doping can stabilize the c-LuH$_3$ phases, which show metallic nature (see Fig. S4 in the Supplementary Materials [38]). To examine the possibility of superconductivity in this phase, we employ a recently developed frozen-phonon method to compute the zone-center electron-phonon coupling (EPC) strength in the present structures. This efficient method can identify the strong EPC candidates in many hydrides because the zone-center EPC shows a strong correlation with the full Brillouin zone EPC in these materials [31]. In Fig. 5, we compute the screened ($\omega$) and unscreened ($\tilde{\omega}$) zone-center phonon frequencies for LuH$_3$ at 1 GPa and 25 GPa, as well as Lu$_8$H$_{21}$N at 1 GPa, respectively. We find the screened and unscreened phonon frequencies are almost the same, resulting in $\lambda_\Gamma \sim 0$ for all these materials. In contrast, most hydride superconductors show $\lambda_\Gamma$ ranging from 0.3 to 1.1 [31]. Therefore, the c-LuH$_3$ phase, regardless of whether it is compressed or N-doped, does not show strong EPC in the zone center.

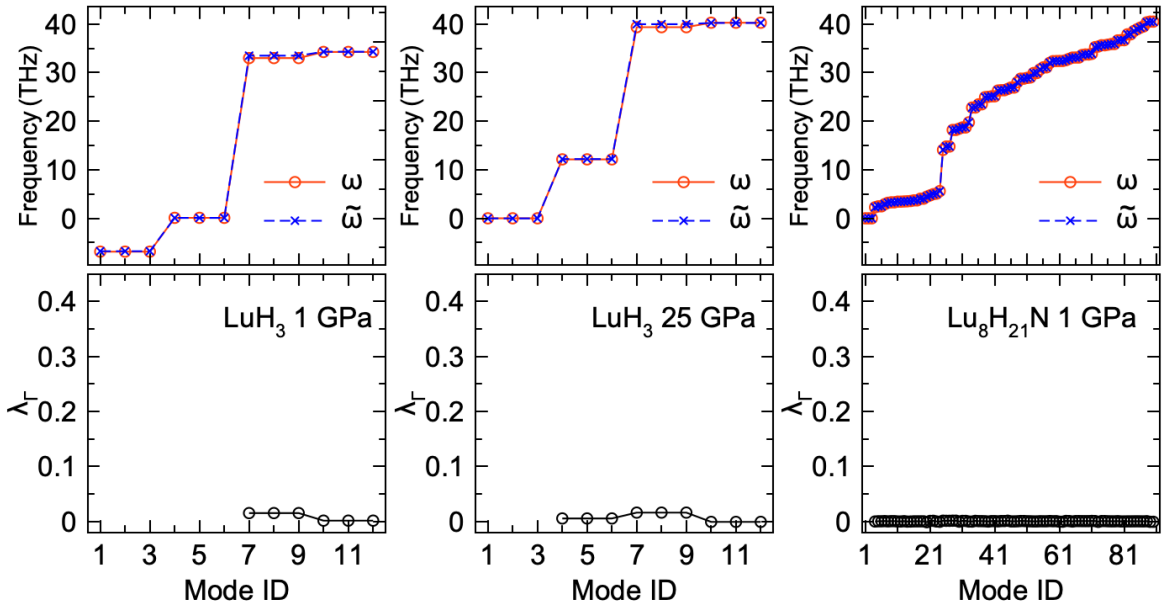

**Fig. 5** Zone-center EPC strength. Upper panels show the screened and unscreened phonon frequencies and lower panels show the zone-center EPC strength.

In summary, we show that both pressure and nitrogen doping can improve the thermodynamic and structural stability of cubic LuH$_3$. A pressure of 25 GPa can stabilize the phonon spectrum of the cubic LuH$_3$ and reduce its enthalpy difference with respect to the stable rhombohedral LuH$_3$ to 12 meV/atom. However, this condition is too far from the recently reported experimental condition in Dasenbrock-Gammon et al. We find that a more realistic way to stabilize the cubic $LuH_3$ is to substitute hydrogen atoms at the tetrahedral sites with nitrogen atoms and introduce vacancies to release the local distortion of hydrogen atoms. This can simultaneously stabilize the phonon spectrum and improve thermodynamic stability. We propose the $Lu_8H_{21}N$ model to address the experimentally synthesized material. In addition, by examining the zone-center electron-phonon coupling, we cannot find any theoretical evidence of strong superconductivity in these structures.